\documentclass[conference]{IEEEtran}

\newcommand{\sav}[1]{{\color{red}Savindu: #1}}

\usepackage{amsmath}
\usepackage[english]{babel}
\usepackage{blindtext}
\usepackage{booktabs} 
\usepackage{array}
\usepackage{multirow}
\usepackage{caption}
\usepackage[FIGBOTCAP]{subfigure}
\usepackage{adjustbox}
\usepackage{enumitem}
\usepackage{colortbl}
\usepackage[normalem]{ulem}
\usepackage{fixltx2e}
\usepackage{soul}
\usepackage{balance}
\usepackage{txfonts}

\newcommand{\ie}{{\em i.e., }}
\newcommand{\eg}{{\em e.g., }}
\newcommand{\myverb}{\fontsize{9}{48}\usefont{OT1}{lmtt}{b}{n}\noindent }
\usepackage{tcolorbox}

\def\BibTeX{{\rm B\kern-.05em{\sc i\kern-.025em b}\kern-.08emT\kern-.1667em\lower.7ex\hbox{E}\kern-.125emX}}

\begin{document}
\title{Unveiling Behavioral Transparency of Protocols Communicated by IoT Networked Assets \\(Full Version)}

\author{
 
 \IEEEauthorblockN{
        Savindu Wannigama\IEEEauthorrefmark{1}, Arunan Sivanathan\IEEEauthorrefmark{3}, Ayyoob Hamza\IEEEauthorrefmark{3} and Hassan Habibi Gharakheili\IEEEauthorrefmark{3}
    }

    \IEEEauthorblockA{
        \IEEEauthorrefmark{1}Department of Computer Engineering, University of Peradeniya, Sri Lanka. Email: \emph{e17369@eng.pdn.ac.lk}
    }
    \IEEEauthorblockA{
        \IEEEauthorrefmark{3}School of EE\&T, UNSW Sydney, Australia. Emails: \{\emph{a.sivanathan, m.ahamedhamza, h.habibi\}@unsw.edu.au}
    }

}

\maketitle

\begin{abstract}
Behavioral transparency for Internet-of-Things (IoT) networked assets involves two distinct yet interconnected tasks: (a) characterizing device types by discerning the patterns exhibited in their network traffic, and (b) assessing vulnerabilities they introduce to the network. While identifying communication protocols, particularly at the application layer, plays a vital role in effective network management, current methods are, at best, ad-hoc.  
Accurate protocol identification and attribute extraction from packet payloads are crucial for distinguishing devices and discovering vulnerabilities. 
This paper makes three contributions: (1) We process a public dataset to construct specific packet traces pertinent to six standard protocols (TLS, HTTP, DNS, NTP, DHCP, and SSDP) of ten commercial IoT devices. We manually analyze TLS and HTTP flows, highlighting their characteristics, parameters, and adherence to best practices---we make our data publicly available; (2) We develop a common model to describe protocol signatures that help with the systematic analysis of protocols even when communicated through non-standard port numbers; and, (3) We evaluate the efficacy of our data models for the six protocols, which constitute approximately 97\% of our dataset. Our data models, except for SSDP in 0.3\% of Amazon Echo's flows, produce no false positives for protocol detection. We draw insights into how various IoT devices behave across those protocols by applying these models to our IoT traces. 
\end{abstract}

\vspace{2mm}

\vspace{0mm}
\section{Introduction} \label{sec:intro}
The rise of IoT devices like cameras and sensors is increasing the size and complexity of computer networks, expanding their attack surfaces. This shift presents notable challenges and, therefore, demands a systematic approach for tasks like asset identification, vulnerability assessment, and cyber risk management. Traditional isolated methods are impractical at this scale, making machine-processable descriptions crucial \cite{MUD, MUdBrick2022} to automate network management functions.

Organizations such as the IETF and NTIA have launched initiatives for standardized data formats to manage cyber risks for IoT devices. The IETF supports the Manufacturer Usage Description (MUD) standard \cite{MUD}, urging IoT manufacturers to specify device functions on networks.
MUD profile enables tight control and verification of IoT device behavior regardless of the operating environment.
NTIA introduced the SBOM framework \cite{SBOM} for a structured, machine-readable inventory of software components and dependencies, offering comprehensive information and hierarchical interrelationships.

MUD and SBOM, while effective in their respective focuses on network behavior and device-embedded security, complement each other by addressing foundational aspects. However, it is important to recognize that these two standards have limitations. They do not comprehensively encompass a vital element---communication protocols. To achieve comprehensive security, it is crucial for network operators to gain visibility into various protocols (such as HTTP, NTP, and TLS) used by individual connected assets. This includes understanding how these protocols are implemented \cite{mudDTLS2024} and configured on the assets, as well as identifying any potential cyber or operational risks these protocols might introduce to their networks and organizations.
While specialized tools effectively analyze specific protocols (like HTTP, SMTP, DNS) for security purposes (\eg flagging inappropriate/illegal website visits, checking malicious email attachments, or detecting data exfiltration), their narrow focus limits their applicability to emerging and heterogeneous protocols. This lack of generalization hinders their effectiveness in diverse and evolving network environments.
Also, existing network intrusion detection systems \cite{Snort, Suricata} and traffic analyzers \cite{Zeek} frequently rely on transport-layer port numbers to define protocols (\eg TCP/80 for HTTP, UDP/123 for NTP). They might also utilize ``proprietary'' signatures \cite{CCS2003} to inspect the content of certain protocols. While metadata like IP protocols, port numbers, or Ether type can aid in protocol identification, they have limitations. They can change over time or even be exploited by malicious applications. Moreover, they often lack a comprehensive set of characteristics unique to each protocol.

This paper aims to set the foundation for systematically specifying and analyzing the characteristics and security parameters of various communication protocols. Our contributions are threefold. \textbf{(1)} We manually analyze packet traces of ten commercial IoT device types, focusing on two representative standard protocols (TLS, HTTP) and highlighting device-specific fingerprints and vulnerabilities in the parameters of these protocols (\S\ref{sec:tool}); 
\textbf{(2)} We develop a common model to describe signatures that help with the systematic analysis of protocols even communicated via non-standard port numbers (\S\ref{sec:map}); 
\textbf{(3)} We evaluate the efficacy and demonstrate the utility of our data models for six protocols (TLS, HTTP, DNS, NTP, DHCP, and SSDP), representing 97\% of IoT network behaviors in our traffic traces, and draw insights into violations of security best-practices across protocols and device types (\S\ref{newInsights}). We publicly release \cite{GitHub2024} our traffic traces and data models.

\section{Attributes and Vulnerabilities of \\Standard Protocols used by IoT Devices}\label{sec:tool}

We obtained packet-level raw traffic traces from a prominent public dataset \cite{18tmc} (widely used by various IoT traffic analytics researchers) to study how IoT devices utilize different communication protocols. This paper focuses on ten IoT device types (\ie Awair air quality, Ring doorbell, Triby speaker, Withings sleep sensor, TP-Link camera, Amazon Echo, Pixtar photo frame, LIFX lightbulb, Samsung camera, and Withings baby monitor) that are fairly popular in the consumer market.    
We concentrate on six standard protocols: TLS, HTTP, DNS, NTP, DHCP, and SSDP. Together, these protocols represent an average of 97\% of flows for each IoT device type within our dataset. In this section, our baseline analysis centers on TLS and HTTP---we will examine all six protocols in \S\ref{newInsights}.

We start by filtering packets of each protocol by employing the standard transport-layer protocol and port numbers\footnote{We acknowledge that our analysis in this section may miss some packets of the intended protocols that occur over non-standard port numbers.} often used for communications: {\myverb{TCP/443}} for TLS (HTTPS), {\myverb{TCP/80}} for HTTP. We manually analyze attributes of each of these two baseline protocols on a per-flow basis, where a unique 5-tuple (source IP address, destination IP address, transport-layer protocol, source port number, and destination port number) identifies each flow. We release our processed per-flow packet traces of the ten IoT devices (studied in this paper) as open data \cite{GitHub2024} for the research community to use. Table~\ref{tab:dataset} summarizes our dataset consisting of about 16M packets and 250K flows---TLS and HTTP combined contribute to a third of these flows. 

For the rest of this section, we manually extract attributes of our chosen two protocols and draw insights into whether and how those attributes can constitute fingerprint patterns to characterize behaviors of various IoT device types. We also highlight vulnerabilities\footnote{The dataset we study in this paper was originally collected in 2016 \cite{18tmc}. The vulnerabilities highlighted here may (or may not) have been remediated by respective IoT device manufacturers or service providers.} in device behaviors, where attribute values of protocols do not conform to corresponding best-practice references (expected settings and/or configurations).  

\subsection{Transport Layer Security (TLS)}\label{sec:tls}

The TLS protocol is typically implemented on top of TCP, enabling a secure end-to-end channel between the client and server to encrypt the contents of applications such as HTTP, FTP, SMTP, and IMAP. 

The TLS handshake consists of three main phases: exchanging keys, exchanging server parameters, and authenticating the server. During the key exchange phase, the client sends a ClientHello message containing (a) ``client-hello-version''; 
(b) ``client-hello-cipher-suites'';  
and (c) ``client-hello-extensions''.  
After the server receives and processes the ClientHello and determines the appropriate cryptographic parameters, it responds with a ServerHello message, containing (d) ``server-hello-version''; the selected TLS version by the server, which is equal to or lower than that suggested by the client in the ``client-hello-version'',  (e) ``server-hello-cipher-suites''; the single cipher suite selected by the server from the list in the ``client-hello-cipher-suites'', and (f) ``server-hello-certificate'', which is a list of extensions selected from ``client-hello-extensions''. The combination of ClientHello and ServerHello determines the shared keys for secure communication \cite{rfc5246}. 

It can be seen from the fourth column of Table~\ref{tab:dataset} that seven of ten IoT devices used the TLS protocol via the standard transport-layer port {\myverb{TCP/443}} across 20,860 flows---TP-Link camera, LIFX lightbulb, and Withings baby monitor had none. 
In what follows, we extract the six attributes discussed above from these TLS flows to conduct our analysis.

\begin{table}[t!]
    \centering
    \caption{Our dataset contains per-flow packet traces of six representative protocols (of which two were analyzed manually) across ten IoT devices. Superscripts are defined as follows. ``$^*$'': unique fingerprint, ``$^{\color{teal}+}$'': secure configs,  ``$^{\color{red}-}$'': vulnerable configs, and ``$^{\color{blue}\dag}$'': not very secure configs.} 
    \label{tab:dataset}
    \vspace{-1mm}
	\renewcommand{\arraystretch}{1.1}
	\begin{adjustbox}{width=0.475\textwidth}
        \begin{tabular}{|l|l|l|l|l|l|l|l|l|}
        \hline
        \multirow{2}{*}{}               & \textbf{\# packets} & \textbf{\# TOTAL} & \textbf{\# TLS flows} & \textbf{\#  HTTP flows} \\ 
                                        &                     &       \textbf{flows}                  & ({\myverb{TCP/443}})               & ({\myverb{TCP/80}})               \\ \hline
        \textbf{Awair air   quality}     & 906,079                    &  2,153                       & 647$^{*{\color{teal}+}{\color{red}-}}$ [+179]                      & 0                                              \\ \hline
        \textbf{Ring doorbell}          & 39,107                    & 633                        & 103$^{*{\color{teal}+}{\color{red}-}}$                       & 3$^{*{\color{teal}+}{\color{red}-}}$                      \\ \hline
        \textbf{Triby speaker}          & 943,513                    & 7,953                        & 3,612$^{*{\color{teal}+}{\color{blue}\dag}}$                      &  187$^{*{\color{red}-}{\color{red}-}}$                       \\ \hline
        \textbf{Withing sleep   sensor} & 685,844                    & 38,899                        & 4,427$^{*{\color{teal}+}{\color{red}-}}$                      & 15,575$^{*{\color{red}-}}$ [+57]                       \\ \hline
        \textbf{TP-Link camera}          & 435,198                    & 10,549                        & 0 [+26]                     & 3,530$^{*{\color{red}-}}$ [+6]                       \\ \hline
        \textbf{Amazon Echo}            & 4,666,750                    & 106,964                        & 10,957$^{*{\color{red}-}{\color{blue}\dag}}$                      &  3,707$^{*{\color{teal}+}}$ [+286]                         \\ \hline
        \textbf{Pixtar photo frame}   & 83,832                    & 6,239                        & 1,049$^{*{\color{red}-}{\color{red}-}}$                     & 23$^{*{\color{teal}+}}$ [+286]                       \\ \hline
        \textbf{LIFX lightbulb}         & 1,368,191                    & 7,740                        & 0 [+131]                     & 0                       \\ \hline
        \textbf{Samsung camera}         &  5,880,402                   & 56,478                        & 65$^{*{\color{red}-}{\color{red}-}}$                    & 13,396$^{*{\color{teal}+}}$ [+19,801]                       \\ \hline
        \textbf{Withings baby monitor}         &  685,844                   &  10,114                       & 0                        & 5,037$^{*{\color{teal}+}}$                       \\ \hline
        \end{tabular}
    \end{adjustbox}  
    \vspace{-3mm}
\end{table}


\textbf{TLS Fingerprints:}  
The client's hello version showed consistent communication preferences for various IoT devices, with some exceptions, like the Samsung camera and Amazon Echo offering multiple versions (\ie v{\myverb{1.2}} to some servers and v{\myverb{1.0}} to others). The server's hello version matched the intended version for most devices except the Pixtar photo frame. Identifiable patterns existed in transmitted ``hello-version" data, but these patterns lacked uniqueness.

\begin{figure}[t!]
    \begin{center}  
       \hspace{-6mm}
        \mbox{
            \subfigure[Ring doorbell and Awair air qlty.]{
                \label{fig:ringawair-cs}
                {\includegraphics[width=0.23\textwidth,height=0.145\textwidth]{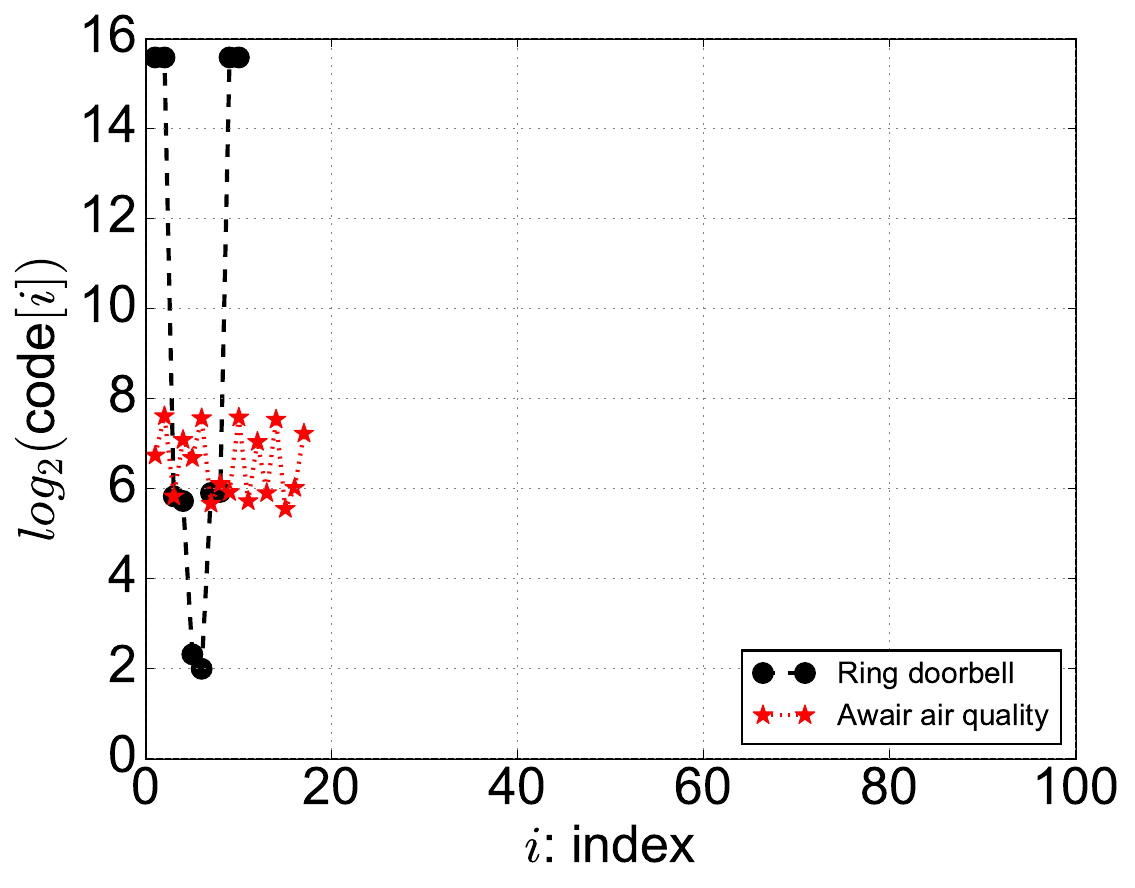}}\quad
            }
        }
        \hspace{-8mm}
        \mbox{
            \subfigure[Amazon Echo.]{
                \label{fig:amazonecho-cs}
                {\includegraphics[width=0.23\textwidth,height=0.145\textwidth]{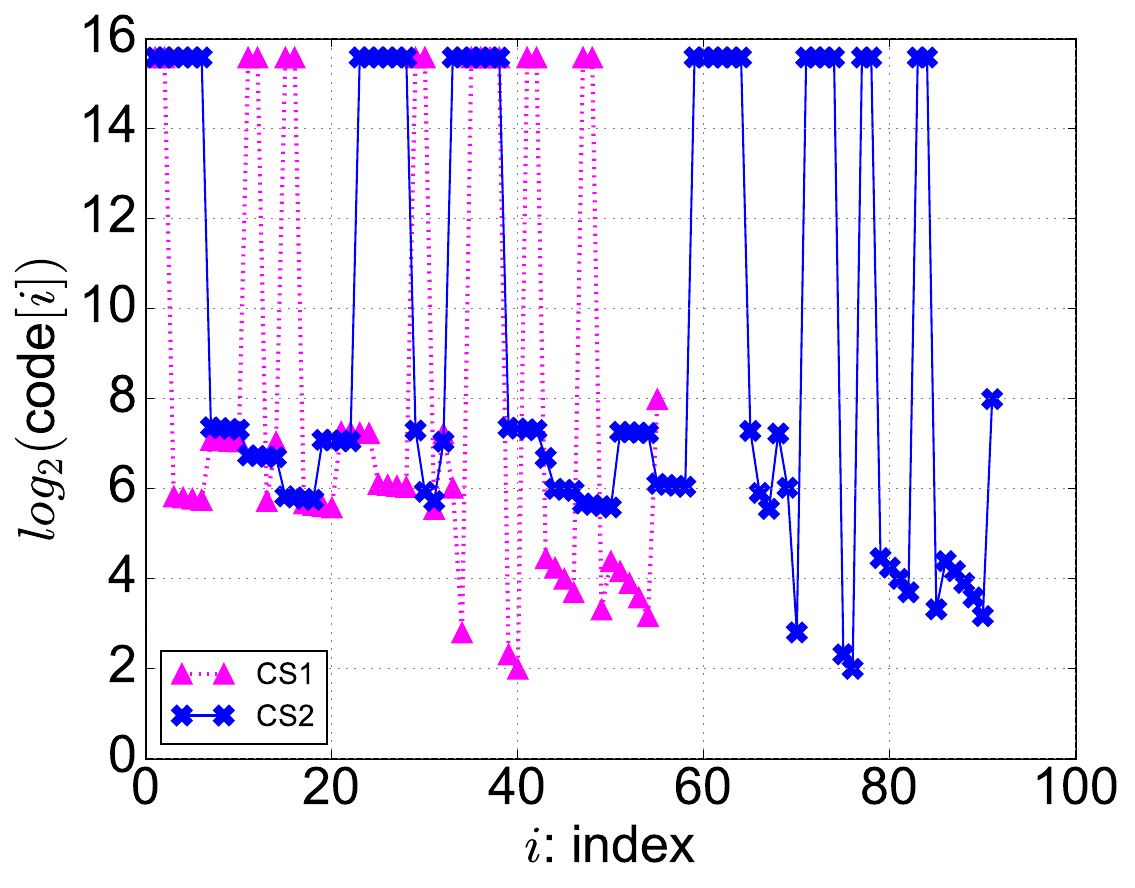}}\quad
            }
        }
        \vspace{-3mm}
        \caption{Ciphersuite fingerprint of: (a) Ring doorbell and Awair air quality, and (b) Amazon Echo.}  
        \label{fig:cs}
        \vspace{-8mm}
    \end{center}
\end{figure}

Our analysis showed that two attributes, namely the client ciphersuite and extensions, display unique fingerprints across the seven IoT device types, which had TLS communicated over {\myverb{TCP/443}} flows.
The ciphersuite is an ordered list of ciphers\footnote{Each cipher is identified by a 4-character hex code. For example, code {\myverb{0xc014}} indicates {\myverb{TLS\_ECDHE\_RSA\_WITH\_AES\_256\_CBC\_SHA}}  \cite{BroadcomCS}, which is offered by the Ring doorbell as its first preferred option.} the client supports. Our analysis revealed that each device distinguishes itself from others by the unique list(s) they provide to their TLS servers---hence the ``$^*$'' superscript in TLS cells of Table~\ref{tab:dataset}. 
Fig.~\ref{fig:cs} illustrates this fingerprint for three 
representative IoT devices.
The x-axis of these plots represents the index (order) of the offered list, while the y-axis denotes the binary logarithm of the decimal equivalent of the hex codes provided in the list---code {\myverb{0xc014}} is denoted by {\myverb{15.59}} in Fig.~\ref{fig:ringawair-cs}. It can be seen that devices like the Ring doorbell and Awair air quality (Fig.~\ref{fig:ringawair-cs}) use a single list for all of their TLS communications, while Amazon Echo is found to offer two different lists (Fig.~\ref{fig:amazonecho-cs}) depending on the TLS server it communicates with. Also, the list of the Ring doorbell was the shortest among the seven devices, consisting of only ten cipher codes, shown by dashed black lines with ``$\medbullet$'' markers in Fig.~\ref{fig:ringawair-cs}. In contrast, Amazon Echo offers the longest list with 91 ciphers, shown by solid blue lines with ``$\times$'' markers in Fig.~\ref{fig:amazonecho-cs}. 
In response to these client offers, the respective servers selected a distinct code in TLS flows of five devices, namely Awair air quality (\ie {\myverb{0x003c}} and {\myverb{0x003d}}), Pixtar photo frame (\ie {\myverb{0x0039}}), Ring doorbell (\ie {\myverb{0xc027}}), Samsung camera (\ie {\myverb{0xc013}}), and Amazon Echo (\ie {\myverb{0x0035}} and {\myverb{0x002f}}).
Amazon Echo's cloud servers select cipher codes shared with other IoT devices: {\myverb{0xc030}} with the Triby speaker and {\myverb{0xc02f}} with the Withings sleep sensor.
Note that server-selected cipher codes may not always be unique across IoT types but can be incorporated into the device fingerprint---becoming a tie-breaker when multiple devices share a list of client-offered ciphersuites.

Considering the client extensions, an optional set of parameters allow for a dynamic negotiation between the client and the server to optimize the TLS connection based on their capabilities and preferences. We found that the Awair air quality does not use any client hello extensions, the Triby speaker uses two unique extensions, Amazon Echo uses eight extensions, and the remaining four devices (with TLS flows) use a single client hello extension. The key takeaway is that this attribute (for those devices that use it) is a unique byte pattern (\eg 134 bytes for Pixtar photo frame versus 73 bytes for Samsung camera, or 307 bytes and 335 bytes for the two different extensions used by Triby speaker), and therefore can be used as the traffic fingerprint of those IoT device types.

\textbf{TLS Vulnerabilities:} Let us now construct a set of reference configurations (obtained from cybersecurity best practices and guidelines) for the TLS protocol.
A measured attribute not matching the reference configuration indicates a vulnerability. The TLS version is a key configuration. 
TLS versions 1.0 and 1.1 were deprecated by the IETF in 2021 \cite{rfc8996}, to modernize platforms and improve security and reliability, we expect higher versions (\ie 1.2 or 1.3) to be offered and selected by client devices and their respective servers. 
Currently, TLS cipher suites are categorized into four groups \cite{infoCS}: ``insecure'' (92 codes), ``weak'' (196 codes), ``secure'' (32 codes), and ``recommended'' (28 codes). This paper expects no ``insecure'' or ``weak'' code to be suggested/selected by the client/server---only ``recommended'' and/or ``secure'' codes are accepted.

From the TLS version perspective, we see a healthy (invulnerable) cluster of devices, including Awair air quality, Ring doorbell, Triby speaker, and Withings sleep sensor. They always offer a recommended TLS version 1.2, which their cloud servers accept---hence the first ``$^{\color{teal}+}$'' in corresponding cells of Table~\ref{tab:dataset}. For a relatively vulnerable cluster, we have a device like Pixtar photo frame, which offers the recommended v1.2. However, its server does not seem conformant to reference best practices and chooses TLS v1.0 for communication---hence the first ``$^{\color{red}-}$'' in corresponding cells. Lastly, the most vulnerable cluster consists of devices like the Samsung camera and Amazon Echo, which consistently (at least with certain servers) initiate TLS handshakes presenting the deprecated v1.0 in their client hello message, leading to insecure TLS communications.

{\color{black}
Analyzing the cipher suites revealed that every device we studied displays a vulnerability, meaning they suggest at least a weak and/or insecure cipher code.
That said, servers behave differently across IoT device types. Let us start with those that select a secure or recommended code from the client-suggested list. The Triby speaker displayed two cipher suites in its TLS fingerprint: one containing 66 codes, of which 46 are weak, and another containing 80 codes, of which 71 are weak. All servers the speaker communicated with consistently selected a secure code {\myverb{0xc030}}.
Note that servers may not always select an invulnerable code---the onus is on the devices to avoid vulnerable codes. 
For example, the intended TLS servers of IoT devices such as Awair air quality, Pixtar photo frame, Ring doorbell, Withings sleep Sensor, and Samsung camera
consistently chose a weak code from the client-suggested ciphersuite. We also see a mixed pattern of vulnerability in the TLS traffic of Amazon Echo. As we saw in Fig~\ref{fig:amazonecho-cs}, this device presents two lists: one with 55 codes, of which 43 are weak and 11 are insecure, and another one with 91 codes, of which 71 are weak and 11 are insecure. In our dataset, some of the Echo servers selected weak codes (\ie {\myverb{0x0035}} and {\myverb{0x002f}}) and others selected secure codes (\ie {\myverb{0xc030}} and {\myverb{0xc02f}}). 
}. 

\vspace{-2mm}
\subsection{Hypertext Transfer Protocol (HTTP)}\label{sec:http}
\vspace{-1mm}

HTTP is a standard client-server protocol used for distributed hypermedia systems. It is versatile, stateless, and extends beyond hypertext for tasks like object management \cite{rfc9112}. 
HTTP requests come with seven attributes, including: (a) ``authentication-credential'', (b) ``authentication-type'', (c) ``request-version'' allows the sender to indicate the format of a message, (d) ``uri'' (uniform resource identifier), which identifies the resource to which a request is applied, (e) ``method'' specifies the action to be performed on the URI, (f) ``host'' specifies the Internet host and port number of the resource being requested, (g) ``user-agent'' of the client initiating a request (\eg browsers, editors, spiders, or web-traversing robots). 
Similarly, HTTP responses contain four attributes, including: (a) ``server'' indicates the application program that accepts connections, (b) ``response-version'' which is similar to the request version above, (c) ``status-code'' is a 3-digit integer code that helps the client better understand the health and/or actions of the server, and (d) ``response phrase'' provides a short description of the status-code.

Among the ten IoT devices in our dataset, eight utilized the HTTP protocol through the standard transport-layer port {\myverb{TCP/80}}, resulting in a total of 41,458 flows, as shown in the fifth column of Table~\ref{tab:dataset}. Awair air quality and LIFX lightbulb were exceptions, with no instances of such usage.

\textbf{HTTP Fingerprints:} 
Our analysis revealed that certain IoT devices exhibit a distinct fingerprint based on the specific set of HTTP methods they utilize. For example, 
Amazon Echo differentiates itself from others by utilizing {\myverb{HEAD}} and {\myverb{PUT}} methods alongside {\myverb{GET}}. Meanwhile, the Triby speaker's distinctiveness lies in its consistent utilization of the {\myverb{GET}} method across all its HTTP requests. In contrast, both the Ring doorbell and Withings baby monitor exclusively utilize the {\myverb{POST}} method for all their HTTP flows. 
Moving to the host attribute, all eight devices communicating HTTP have at least one host exclusive to each of them. Interestingly, even devices from the same manufacturer (\ie Withings sleep sensor and Withings baby monitor) display their individual unique hosts. We also found that the URI attribute bears a distinctive fingerprint for each type of IoT device.

All eight devices consistently use HTTP/1.1. Notably, the Withings sleep sensor, Triby speaker, and TP-Link camera utilize both v1.0 and v1.1 with different servers. Authentication attributes were found only in the HTTP traffic of the Ring doorbell and Triby speaker. These devices use Basic authentication over TCP/80 flows.

Lastly, regarding user agents, all devices except the Withings baby monitor employ more than one agent. The baby monitor exclusively uses ``{\myverb{Withings UserAgent}}'' for all its HTTP flows. In contrast, the sleep sensor (from the same manufacturer) adopts this user agent along with three others.  We identified distinct user agents like ``{\myverb{curl}}'' utilized by the Triby speaker, ``{\myverb{kindle/2.0}}'' by the Amazon Echo, ``{\myverb{Spotify}}'' by the Withings sleep sensor, and ``{\myverb{Microsoft-Windows}}'' by the Samsung camera. 
Devices exhibited shared user agents: ``{\myverb{Mozilla/5.0}}'' among Amazon Echo, Pixtar photo frame, TP-Link camera, and Samsung camera; and ``{\myverb{Dalvik/2.1.0}}'' among TP-Link camera, Triby speaker, and Samsung camera.

Moving to HTTP responses, certain devices consistently connect with a single server via all their HTTP communications, whereas others engage with multiple servers. In the category of devices connecting to a single server, we identified two IoT device types: the Ring doorbell exclusively connects to ``{\myverb{nginx}}'', while the Withings baby monitor solely utilizes ``{\myverb{Apache}}''. However, we note that these two servers cannot be considered unique fingerprints, as the {\myverb{nginx}} application server is shared by other four devices (\ie Pixtar photo frame, Triby speaker, Amazon Echo, and Withings sleep sensor), and the {\myverb{Apache}} application server is used by other five devices (\ie TP-Link camera, Triby speaker, Amazon Echo, Withings sleep sensor, and Samsung camera).
In the second category, where multiple servers are involved, five devices stand out for their distinct server(s): ``{\myverb{relayd}}'' and ``{\myverb{lighttpd}}'' for TP-Link camera, ``{\myverb{AkamaiNetStorage}}'', ``{\myverb{Microsoft-IIS}}'' and ``{\myverb{AmazonS3}}'' for Amazon Echo, ``{\myverb{cloudflare-nginx}}'' for Withings sleep sensor, ``{\myverb{SmartCamWebService}}'', and ``{\myverb{AkamaiGHost}}'' for Samsung camera. Note that the corresponding devices use these distinct servers less frequently than shared servers.

Examining the status code, as expected, ``{\myverb{200 OK}}'' is prevalent among all eight devices. Surprisingly,
the Triby speaker is found to frequently receive ``{\myverb{206 Partial Content}}'' in over half of its HTTP flows. We also identified other distinctive status codes: ``{\myverb{400 Bad Request}}'' for Amazon Echo and ``{\myverb{503 Service Unavailable}}'' for Samsung camera---but relatively rare, appearing in less than 1\% of the total HTTP flows of respective devices.       

\underline{Takeaway:} Despite overlaps among device types in certain attributes, combining multiple HTTP attributes will indeed lead to a distinct and robust identifying pattern, highlighted by ``$^{*}$'' in corresponding cells under HTTP flows in Table~\ref{tab:dataset}.

\textbf{HTTP Vulnerabilities:} For baseline security best practices, we prioritize two essential HTTP attributes: version and authentication. HTTP has five versions: v0.9, v1.0, v1.1, v2.0, v3.0. 
RFC7230 \cite{rfc7230} has deprecated HTTP/0.9 due to its lack of support for request header fields, rendering it incompatible with name-based virtual hosts. Furthermore, HTTP/1.0 inadequately addresses the impacts of hierarchical proxies, caching, the necessity for persistent connections, and the use of name-based virtual hosts \cite{rfc7230}. Consequently, adopting HTTP/1.1 or a more recent version is deemed a recommended practice. Moving to the second essential attribute, Basic authentication is considered insecure \cite{PARVP2021} compared to other options (\eg Digest). This is due to the practice of transmitting user-identifying information and passwords over the network in plaintext \cite{rfc7617}, which poses significant security risks. Digest, on the other hand, relies on cryptographic hashes \cite{rfc2617}.

Regarding the HTTP version, there are three devices---TP-Link camera, Triby speaker, and Withings sleep sensor---that exhibit vulnerability due to the utilization of v1.0 in instances of HTTP flows, and hence the first ``$^{\color{red}-}$'' in corresponding cells under HTTP flows in Table~\ref{tab:dataset}. Note that these three devices employ a combination of both HTTP/1.0 and HTTP/1.1.
Considering authentication, the Ring doorbell and Triby speaker are the only two devices that use the vulnerable ``Basic'' authentication scheme, as highlighted by the second ``$^{\color{red}-}$'' in corresponding cells under HTTP flows in Table~\ref{tab:dataset}.

In summary, we noted vulnerabilities in the HTTP communications of four devices. However, the Pixtar photo frame, Amazon Echo, Samsung camera, and Withings baby monitor remain unaffected by the vulnerabilities we investigated.

\vspace{-2mm}
\section{Data Models for Protocols}\label{sec:map}
As discussed in the previous section, each protocol operates in a 
specific manner, guided by the specifications of its originating organization. For example, HTTP messages have well-defined semantics, including request methods, request header fields, response status codes, response header fields, and the payload of messages. Furthermore, protocols are often unique \cite{CCS2003}, displaying identifiable patterns in traffic headers, contents, or statistical characteristics. Nonetheless, a significant challenge arises in determining whether and how we can formally specify those patterns, at the packet-level and/or flow-level, inherent in heterogeneous protocols along with their attributes (\eg protocol version, authentication method, credentials, user agent, or negotiated cipher). This challenge pertains to devising a unified schema that can accommodate these diverse elements. Such schema would serve as a valuable enhancement to the existing MUD standard \cite{MUD} for IoT devices by offering supplementary insights into the specific protocols spoken through individual flows or access control entries (ACEs) within the MUD file.

\begin{figure}[t!]
	\centering
	\includegraphics[width=0.955\linewidth]{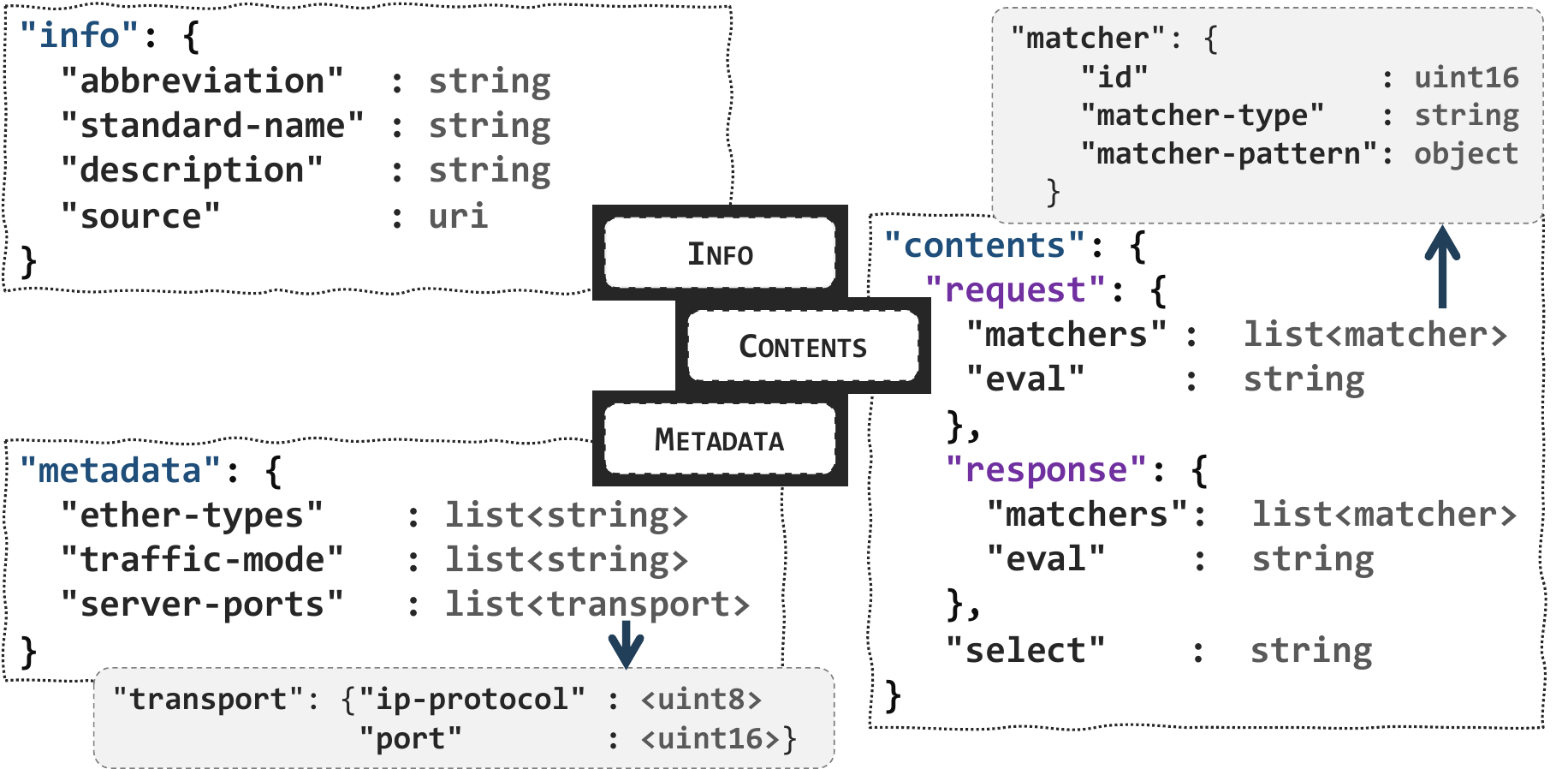}
	\vspace{-2mm}
	\caption{Visual representation of our protocol data schema.}\label{fig:schema}
    \vspace{-7mm}
\end{figure}

\textbf{Protocol Schema:} We envision a minimal structure comprising three central sections: \textsc{Info}, \textsc{Metadata}, and \textsc{Contents}, shown in Fig.~\ref{fig:schema}.
This model encapsulates characteristics of a group of packets associated with a ``protocol flow'' consisting of a couple of unidirectional flows (request and response of a given 5-tuple). Note that the schema is expected to accommodate partial characterizations, meaning certain sections and/or fields may be absent. Our schema could be extended to incorporate protocol parameters (attributes)---such extension is beyond the scope of this work and left for future research.   

The \textsc{Info} section (top left in Fig.~\ref{fig:schema}) provides high-level information about the data model. At a minimum, it should contain fields such as ``{\myverb{abbreviation}}'' (\eg HTTP), ``{\myverb{standard-name}}'' (\eg hyper text transfer protocol), and ``{\myverb{description}}''. Also, ``{\myverb{source}}'' refers to a public source of information (\eg an RFC) to be used for the subject protocol. 

The \textsc{Metadata} section (bottom left in Fig.~\ref{fig:schema}) captures information mainly related to the network flow a protocol uses for communication. Three fields are dedicated to describing headers of the data-link layer (``{\myverb{ether-types}}''), network layer (``{\myverb{traffic-mode}}'', \eg unicast), and transport layer (``{\myverb{server-ports}}''). Ports can be a list of \{{\myverb{ip-protocol}},{\myverb{port}}\} like \{TCP/80, TCP/8008, TCP/8080, TCP/8888\} for HTTP.

The \textsc{Contents} section (right in Fig.~\ref{fig:schema}) describes patterns (strings \cite{regexAPI}, bytes \cite{byteseekAPI}, or statistical measures) in the content of packets exchanged either in ``{\myverb{request}}'', ``{\myverb{response}}'', or ``{\myverb{combined}}'' directions. Depending on the complexity of a given communication protocol, a list of different patterns (\eg string, byte, or statistical distribution) may exist in its packet contents. 
Obviously, various patterns would require their own ``{\myverb{matchers}}''. As it can be seen at the top right of Fig.~\ref{fig:schema}, each matcher comes with an ``{\myverb{id}}'', a ``{\myverb{matcher-type}}'' (\eg string, byte, statistical), and a corresponding ``{\myverb{mactcher-pattern}}''. The field ``{\myverb{eval}}'' is a Boolean expression of matcher IDs, indicating how various matchers (if any) are jointly applied. Lastly, the ``{\myverb{select}}'' field is a Boolean expression on how signatures of request, response, and/or combined directions are incorporated to yield the final inference outcome.  

\textbf{Benefits:}
The protocol model provides essential insights into device communication patterns, benefiting manufacturers/developers by enhancing the selection and inherent parameters of protocols for the devices they produce. Network operators can leverage it to reduce attack surfaces by adjusting configurations on their network, or develop enforceable security policies across network segments. One may also use protocol data models for anomaly detection purposes.

\textbf{How to Use:} 
These models will ultimately be provided by the community (\eg protocol developers or researchers) in an open and transparent manner. Protocol models can be utilized in conjunction with MUD files through passive network monitoring tools to filter traffic of intended devices across specific flows (\eg by  5-tuple), facilitating real-time and scalable protocol detection and behavior characterization. 

\textbf{Our Usecase:}
The protocol data models are designed to offer advanced and systematic computing capabilities for managing network-level security. They facilitate the detection of protocols beyond port numbers, enabling vulnerability assessment and automatic compliance checks against best practices. 


\vspace{-1mm}
\section{Realizing Protocol Data Models: \\Evaluation and Utility} \label{newInsights}

\begin{figure*}[h!]
    \begin{center}  
        \mbox{
            \subfigure[HTTP.]{
                \label{fig:http-content}
                {\includegraphics[width=0.37\textwidth]{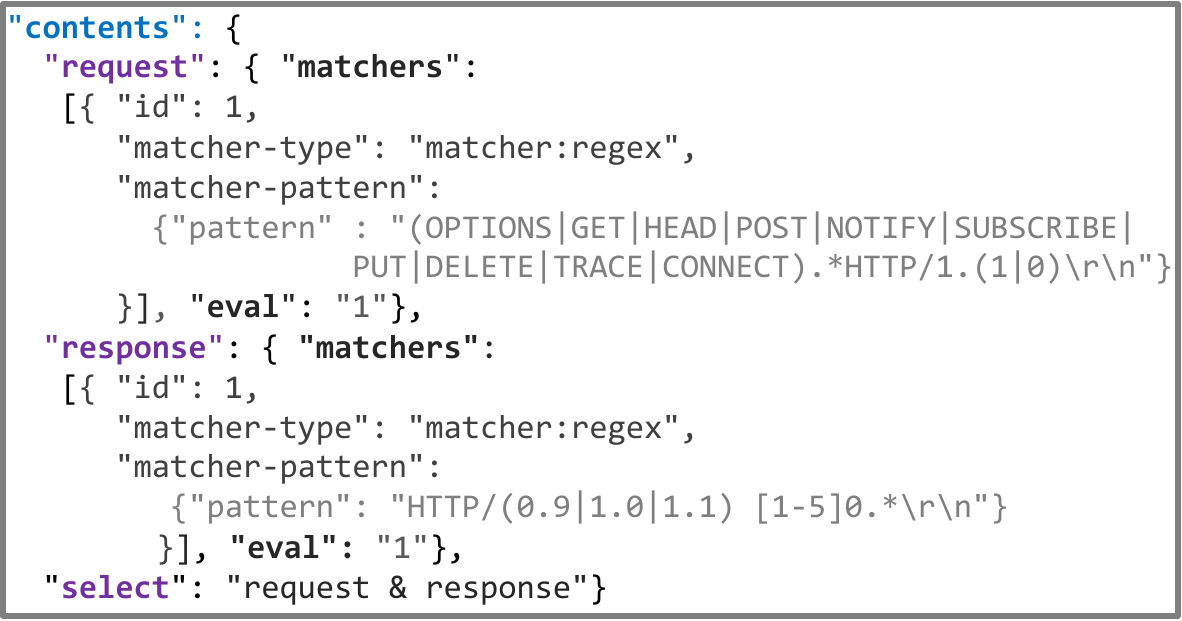}}\quad
            }
        }
        \hspace{-4mm}
        \mbox{
            \subfigure[TLS.]{  
                \label{fig:tls-content}
                {\includegraphics[width=0.37\textwidth]{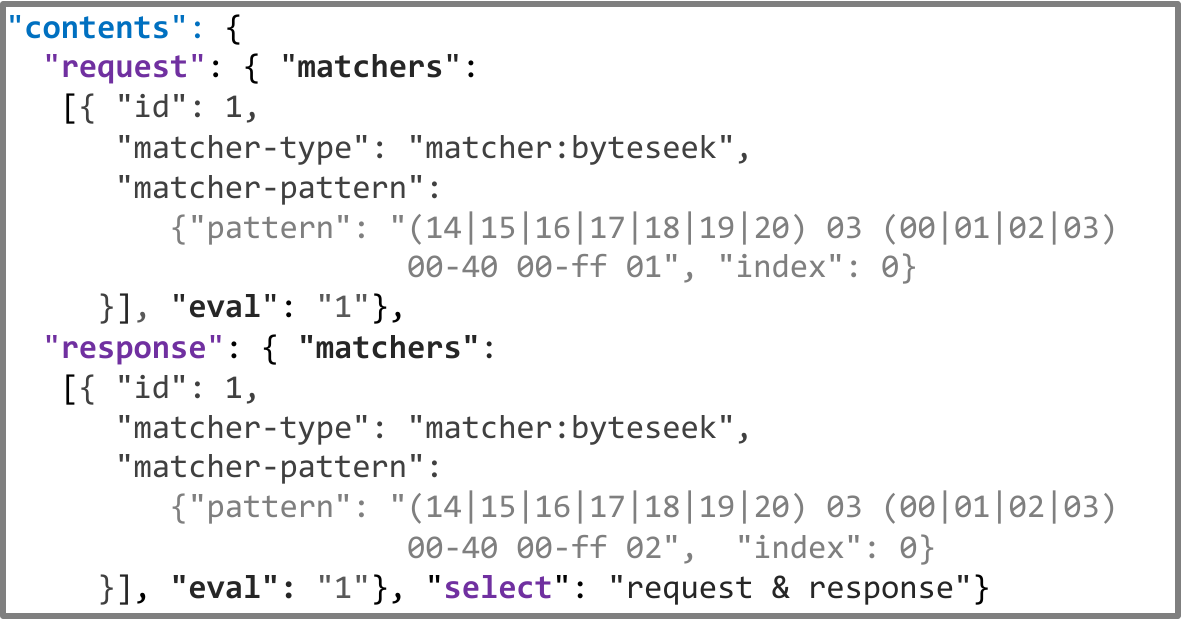}}\quad
            }
        }
        \vspace{-3mm}
        \caption{The \textsc{Contents} section of our protocol data model: (a) HTTP and (b) TLS.}  
        \label{fig:data-models}
        \vspace{-7mm}
    \end{center}
\end{figure*}

Through manual analysis outlined in \S\ref{sec:tool} and drawing inspiration from previous studies \cite{CCS2003,USENIX2006}, we created the data model for six commonly used IoT protocols specifically chosen for this paper: TLS, HTTP, DNS, NTP, DHCP, and SSDP.
In the cases of HTTP and SSDP, content signatures are represented by regular expressions---an example is shown in Fig.~\ref{fig:http-content}. In contrast, for TLS, DNS, NTP, and DHCP, the content structure is characterized by byte patterns---an example is shown in Fig.~\ref{fig:tls-content}. 

We iterate each of the six data models to all flows of individual devices in our dataset---instead of solely considering the specific transport-layer port numbers---to isolate flows that match the subject protocol. Once a protocol is detected, we employ manual analysis to examine its attributes---this step can be automated by extending our schema---looking for potential violations in adherence to security best practices.
Our manual verification confirmed that, except for the SSDP model in 293 flows ($0.3$\%) of Amazon Echo\footnote{All of these unsolicited incoming UDP flows were found to 
consist of a single packet, lacking the expected attributes for the SSDP protocol.}, our data models produced no false positives in detecting the protocols studied in this paper. Recall that the six protocols we analyzed represent $97$\% of IoT behavior (all flows per device in our dataset).
In what follows, we start by analyzing new results derived using our data models for TLS and HTTP and comparing them with those from our analysis in \S\ref{sec:tool}. Next, we analyze the protocol characteristics and vulnerabilities of DNS, NTP, DHCP, and SSDP. 


\subsection{TLS and HTTP: New Findings}
The data models facilitated the automatic discovery of previously unnoticed TLS and HTTP flows, in addition to those that occurred on standard port numbers (\S\ref{sec:tool}). The counts of newly detected flows are indicated within square brackets in Table~\ref{tab:dataset}. 
Note that the LIFX lightbulb and TP-Link camera, previously thought to lack TLS flows, are discovered to engage in TLS communication on non-standard port numbers. Specifically, the lightbulb has 131 TLS flows across port {\myverb{TCP/56700}}, while the camera employs TLS for 26 flows over {\myverb{TCP/50443}}. Furthermore, we identified the Awair air quality monitor to have an additional 179 TLS flows via a dedicated service port, {\myverb{TCP/8883}}.
Moving to HTTP, we found five devices communicating this protocol over non-standard ports. The Withings sleep sensor communicated 57 HTTP flows over {\myverb{TCP/7888}}; TP-Link camera revealed {\myverb{TCP/8080}} in 6 flows; Amazon Echo communicates HTTP over two non-standard ports: {\myverb{TCP/49153}} with 88 flows and {\myverb{TCP/49154}} with 198 flows; the Pixtar photo frame exhibited three new ports in its HTTP communications: {\myverb{TCP/49152}} with 166 flows, {\myverb{TCP/5000}} with 60 flows, and {\myverb{TCP/46380}} with 60 flows; and, the Samsung camera uses HTTP over {\myverb{TCP/49152}} across 19,801 flows.

We applied TLS and HTTP best-practice references in \S\ref{sec:tls} and \S\ref{sec:http} to the new flows discovered in this section. 
LIFX lightbulb and TP-Link camera are found to be vulnerable in all TLS flows due to old (client/server) versions.   
The lightbulb offered three cipher codes: two weak and one insecure. 
On the other hand, the camera offered 21 codes to its servers, of which 10 are weak, and another 10 are insecure. For both cases, servers always chose the weak code {\myverb{0x002F}}.
Lastly, the HTTP data model revealed that the TP-Link camera uses Basic authentication in two newly discovered HTTP flows---hence, a sign of potential vulnerability. 
The new findings demonstrate the superiority of data models over traditional port-based methods for protocol identification and characterization.

\subsection{Characteristics of DNS, NTP, DHCP, and SSDP}
In this section, we employ data models to gain insights into how various devices use these four protocols, demonstrating their efficacy in security management. Table~\ref{tab:dataMdlRes} summarizes the results for these four protocols.

\textbf{DNS:} All ten IoT devices whose traffic we examined used DNS in 140,156 flows via {\myverb{UDP/53}} to obtain the IP address of their intended cloud servers. Our data models confirmed the consistent use of DNS protocol among all IoT devices through the standard port, and no vulnerabilities were detected regarding DNS, indicating adherence to best practices.


\textbf{NTP:} This protocol serves the purpose of synchronizing computer clocks on the Internet \cite{rfc5905}. We observed that seven out of ten IoT devices in our dataset use NTP in their network traffic: Awair air quality, LIFX lightbulb, Ring doorbell, TP-Link camera, Triby speaker, Amazon Echo, and Samsung camera, with a total 18,207 NTP flows---all communicated over the standard {\myverb{UDP/123}}. 
When considering attribute vulnerabilities, adhering to current best practices (RFC8633 \cite{rfc8633}) entails avoiding the use of NTP versions older than v4 to mitigate the risk of known attacks. Additionally, RFC8633 emphasizes the exclusive utilization of broadcast and symmetric variants of NTP on trusted networks and with trusted peers---use of ``client-mode=3'' and ``server-mode=4'' over the Internet is recommended as a best practice. 
Lastly, many known attacks on NTP set the origin timestamp (``org'') to zero in NTP response---hence, it is best to avoid it. Our findings indicate that all seven devices, except the Triby speaker, exhibit vulnerabilities stemming from utilizing NTP v3 for either client or server functions. One may consider a vulnerable client version a higher risk factor than the server. Specifically, all devices, except for the Triby speaker and Amazon Echo, employed NTP v3 as their client version. Regarding the server org value, the Ring doorbell is the only vulnerable device---all of its flows come with this risk. 

\begin{table}[t!]
    \centering
    \caption{Protocol flows detected by four protocol models, extending our analysis in Table~\ref{tab:dataset}.} 
    \label{tab:dataMdlRes}
    \vspace{-1mm}
	\renewcommand{\arraystretch}{1.1}
	\begin{adjustbox}{width=0.475\textwidth}
        \begin{tabular}{|l|l|l|l|l|}
        \hline
        \multirow{2}{*}{}               & \textbf{\# DNS flows} & \textbf{\# NTP flows} & \textbf{\# DHCP flows} & \textbf{\# SSDP flows} \\ 
        \hline
        \textbf{Awair air   quality}     & 1,118$^{\color{teal}+}$  & 190$^{{\color{teal}+}{\color{red}-}}$ & 2$^{\color{teal}+}$ & 0 \\ \hline
        \textbf{Ring doorbell}           & 486$^{\color{teal}+}$    & 26$^{{\color{red}-}{\color{red}-}}$  & 2$^{\color{teal}+}$ & 0 \\ \hline
        \textbf{Triby speaker}           & 3,914$^{\color{teal}+}$  & 111$^{{\color{teal}+}{\color{teal}+}}$ & 4$^{\color{teal}+}$ & 0 \\ \hline
        \textbf{Withing sleep   sensor}  & 18,759$^{\color{teal}+}$ & 0  & 6$^{\color{teal}+}$ & 1$^{\color{red}-}$ \\ \hline
        \textbf{TP-Link camera}           & 4,826$^{\color{teal}+}$  & 41$^{{\color{teal}+}{\color{red}-}}$ & 2$^{\color{teal}+}$ & 0 \\ \hline
        \textbf{Amazon Echo}             & 73,613$^{\color{teal}+}$ & 16,293$^{{\color{teal}+}{\color{red}-}}$ & 12$^{\color{teal}+}$ & 15$^{\color{red}-}$ \\ \hline
        \textbf{Pixtar photo frame}      & 4,842$^{\color{teal}+}$  & 0 & 2$^{\color{teal}+}$ & 2$^{\color{teal}+}$ \\ \hline
        \textbf{LIFX lightbulb}          & 6,292$^{\color{teal}+}$  & 1,117$^{{\color{teal}+}{\color{red}-}}$ & 2$^{\color{teal}+}$ & 0 \\ \hline
        \textbf{Samsung camera}          & 21,238$^{\color{teal}+}$ & 429$^{{\color{teal}+}{\color{red}-}}$ & 21$^{\color{teal}+}$ & 687$^{\color{teal}+}$ \\ \hline
        \textbf{Withings baby monitor}   & 5,068$^{\color{teal}+}$  & 0  & 2$^{\color{teal}+}$ & 0 \\ \hline
        \end{tabular}
    \end{adjustbox}  
    \vspace{-5mm}
\end{table}

\textbf{DHCP: }
It is another foundational protocol that allows devices to onboard the network, systematically obtaining network configuration parameters \cite{rfc2131}. 
Note that DHCP traffic is infrequently used only at the time of joining the network. Our dataset only contains 55 DHCP flows, all via the standard port {\myverb{UDP/67}}. While DHCP revealed device-specific fingerprints \cite{23nomsActivePAssive}, we found no vulnerabilities in the detected DHCP flows.

\textbf{SSDP: }
This discovery and search protocol, utilizing a multicast variant of HTTP over UDP \cite{upnp}, is not widely observed in our dataset. Only four devices---Pixtar photo frame, Amazon Echo, Withings sleep sensor, and Samsung camera---utilized SSDP. Applying the SSDP data model yielded a total of 705 flows, all of which were verified to be exclusively communicated via the standard port {\myverb{UDP/1900}}. 
For SSDP best-practice reference, we refer to the OCF guidelines \cite{upnp}: for the notification sub-type attribute the following values are considered acceptable: ``ssdp:alive'', ``ssdp:byebye'', ``ssdp:update'', ``upnp:propchange''; the search-type attribute of M-SEARCH packets should be ``ssdp:discover''; and, the time-to-live (TTL) of each IP packet for each multicast request message should default to 2---TTL larger than 1 extends the discovery range, but introduces susceptibility to DDoS attacks. We found that Amazon Echo (TTL=4) and Withings sleep sensor (TTL=64) do not adhere to SSDP best practices.

\vspace{-1mm}
\section{Related Work}\label{sec:related}

\textbf{IoT Risk Assessment:} 
In recent years, assessing the security posture of IoT devices received growing attention from researchers. Work in \cite{Loi2017-nh} actively subjected consumer IoT devices to automated tests, objectively evaluating metrics like confidentiality, integrity, or access control. PARVP \cite{PARVP2021} took a different approach by developing a passive risk assessment system to ascertain the security of HTTP-based authentication methods used by networked cameras.
IoTLS \cite{iotls} focused on the consumption of TLS by IoT devices and employed both active and passive approaches to highlight vulnerabilities within this protocol. Their investigation covered aspects such as protocol versions, cipher suites, certificate validation, and maintenance of root stores.
Despite some overlap, our work broadens the scope by encompassing five additional protocols. Moreover, our objectives go beyond ad-hoc evaluations. We advocate structured data models that can passively verify best-practice adherence at larger scales.

\textbf{IoT Device Characterization:} 
The characterization of IoT network behaviors has received interest from both academia and industry. Various methods, including active \cite{Avast2019,ICIAfS2018}, passive \cite{Marchal19,18tmc}, and combined \cite{23nomsActivePAssive} approaches, have been employed to classify IoT device behaviors. For instance, AuDI \cite{Marchal19} developed unsupervised machine learning models, relying on time-series signals such as NTP and DNS flows to generate IoT device fingerprints. Authors of \cite{18tmc} utilized both packet-based (\eg server ports, domain names) and flow-based features (\eg statistical flow volume/duration) to predict IoT device make/model from passive measurements. Additionally, work in \cite{MUDcheck2022} utilized MUD files as fingerprints, constructing a graph from real-time IoT device network behavior and applying similarity measures to classify the graph. We aim to advance this research by developing protocol-specific fingerprints for more granular deterministic characterization of IoT device classes, building upon prior relevant work.

\vspace{-1mm}
\section{Conclusion}
IoT devices expand network attack surfaces and often remain unmanaged. Gaining systematic visibility into their behavior is crucial for managing cyber risks. This paper focused on IoT device communication protocols.
We manually examined device-specific fingerprints and vulnerabilities from real traffic traces across two prevalent protocols (TLS, HTTP), used by ten distinct commercial IoT device types. Subsequently, we developed a machine-processable data schema to formally specify protocol signatures, enabling systematic traffic analysis. Finally, we evaluated the data models of six protocols (TLS, HTTP, DNS, NTP, DHCP, and SSDP) by applying them to IoT traces, and highlighted vulnerabilities in protocol parameters on both standard and non-standard ports.

\vspace{-2mm}
\bibliographystyle{IEEEtran}
\bibliography{ProtoIoT}

    \begin{appendices}

    \end{appendices}


    
    
    


\end{document}